\documentclass[twocolumn,showpacs,aps,superscriptaddress]{revtex4}
\usepackage{graphicx}
\usepackage{bm}
\usepackage{amsmath}
\usepackage{amsfonts}
\usepackage{amssymb}
\usepackage[latin1]{inputenc}

\newcommand{\ket}[1]{|#1\rangle}

\newcommand{\scalar}[2]{\langle#1|#2\rangle}

\def\Tr{{\rm Tr}}
\def\sys{{\rm sys}}
\def\Env{{\rm Env}}
\def\Var{{\rm Var}}

\begin{document}

\title{Tunneling effects in a one-dimensional quantum walk}
\author{Mostafa Annabestani}
\email{Annabestani@modares.ac.ir}
\affiliation{Department of Physics, Basic Sciences Faculty, Tarbiat Modares University, Tehran, Iran}
\author{Seyed Javad Akhtarshenas}
\email{akhtarshenas@phys.ui.ac.ir}
\affiliation{Department of
Physics, University of Isfahan, Isfahan, Iran}
\affiliation{Quantum Optics Group, University of Isfahan, Isfahan, Iran}
\author{Mohamad Reza Abolhassani}
\affiliation{Department of Physics, Basic Sciences Faculty, Tarbiat Modares University, Tehran, Iran}

\date{\today}
\begin{abstract}
In this article we investigate the effects of shifting position
decoherence, arisen from the tunneling effect in the experimental
realization of the quantum walk,  on the one-dimensional discreet
time quantum walk.  We show that in the regime of this type of noise
the quantum behavior of the walker does not fade, in contrary to the
coin decoherence for which the walker undergos the
quantum-to-classical transition even for weak  noise. Particularly,
we show that the quadratic dependency of the variance on the time
and also the coin-position entanglement, i.e. two important quantum
aspects  of the coherent quantum walk, are preserved in the presence
of tunneling decoherence. Furthermore, we present an explicit
expression for the probability distribution of  decoherent
one-dimensional quantum walk in terms of the corresponding coherent
probabilities, and show that this type of decoherence smooths the
probability distribution.
\end{abstract}



\maketitle

\section{Introduction}
\label{sec:intro}

The quantum walk (QW) is the quantum version of the classical random
walk (CRW). The superposition property of quantum systems allows us
to define quantum walker which can walking in all possible paths,
simultaneously, and the interference of these paths makes some
important differences between the QW and the CRW. Notable
differences between the QW and the CRW are the quadratic dependency
of the variance on the number of steps and the complex oscillatory
probability distribution in the QW instead of the linear variance
dependency and the binomial probability distribution in the CRW.
These differences have been used to present several quantum
algorithms in order to solve some specific problems
\cite{Shenvi03,Ambainis03,AKR,Ambainis05,Childs04,Tulsi08} with
performances better than the best known classical versions.

Two types of QW have been introduced: \textit{discrete time}
\cite{Aharonov} and \textit{continuous time} \cite{FG98}. Although the
mathematical structure is completely different in these two types,
but both have the same results in our problems. The relation between
these two types of QW was an open problem for several years, but
this relation has been found recently \cite{Childs08,Strauch}. A lot
of works have been devoted to the study of QW, both from theoretical
as well as experimental view points.  Many aspects of QW have been
studied theoretically, for example some researches focus on the
network over which the walks takes place. In this area, the QW on a
line has been well studied
\cite{Nayak,qw-markov,Konno,Venegas,konno-book,Venegas-book}, but
other topologies such as cycles \cite{cycle1,cycle2},
two-dimensional lattices \cite{Mackay,Carneiro,Omar,Konno2},
or $n$-dimensional hypercubes \cite{Moore,Marquezino} have also been
investigated.

The study of quantum entanglement and its connection with the QW is
another important research area which has recently attracted much
attention. The effect of  entanglement of the coin subspace on the
QW \cite{Omar,Venegas,Chaobin}, entanglement between the coin and
the position subspaces \cite{Carneiro,Abal06-ent,annabestani} and
the QW as the entanglement generators
\cite{Goyal,Venegas-entanglementgeneration} are examples of these
studies.

Beside of all theoretical studies, the experimental implementation
and realization of the QW are also received much attention
\cite{Travaglione,Knight-OC,Dur-prop,Du,Ryan,Zhao-prop}. In the
experimental implementation, the environment effects will be so
important because, in practice,  the preparation of pure quantum
states without interaction with the environment is impossible, and
the environment can disturb the quantum states and fades the quantum
properties. Therefore, it is very important to formulate and
quantify the influence of decoherence on the QW and, indeed, several
valuable works have been made in this subject
\cite{Kendon2003,KT03,Brun03,Brun2,Dur-prop,Oliveira,lopez,deco,zhang,annabestani-deco}.
Moreover, the study of decoherent QW can be considered as
 a link between  the theoretical study and the experimental implementation.

It is known that the quantum properties of QW, such as the quadratic
dependency of variance on the time, make the QW more powerful than
the CRW, but these properties are very fragile when they are exposed
to decoherence or noise.  Brun et al. \cite{Brun2} have shown that
the quadratic term of variance vanish, in long time limit, even for
weak decoherence of the coin subspace. On the other hand,  the
linear term remains greater than the classical one for reasonable
strength of noise.  It means that if the noise on the system be less
than some certain value, the QW will spread faster than the CRW.
Since noise on the quantum systems is unavoidable and is, indeed, a
mechanism for emergence of the classical behavior in quantum
systems, some authors have tried to use such classical behavior
emerged by the noise. For example, Kendon et al. \cite{KT03} have
shown that a weak noise, both on the coin and the position
subspaces, can be useful in quantum algorithms. They have explained
that a general weak noise can smooths and flats the probability
distribution. Of course, we should note that this smoothness occurs
in the cost of losing the speed of spreading (variance). On the
other hand, other route to classic behavior of quantum walk has been
investigated and compared to classical behavior emerging by
decoherence. Brun \textit{et al.} \cite{Brun2} have shown that
multiple coins version of QW retains ``quantum'' quadratic growth of
the variance except in the limit of a new coin for every step.

In this paper we use our analytical expression for variance,
recently presented in \cite{annabestani-deco},  and show that the
shifting position decoherence does not change the quadratic behavior
of the variance in the 1DQW. We show also that although this
quadratic term depends on the initial state, but it can not be
removed even for initial mixed states. Furthermore, we show that the
effect of this type of decoherence is to smooth the probability
distribution and we present a closed formula for the probability
distribution in terms of the probabilities of the corresponding
coherent QW.

This work is organized as follows. Section II gives a brief review
on the 1DQW and decoherence.  Our calculations and results for
tunneling effects have been put in section III. In this section we
obtain an expression for the variance of 1DQW in the presence of
tunneling noise and show that the time quadratic dependency of the
variance exists for all initial states.  The entanglement properties
between the coin and the position degrees of freedom have been also
studied in section III, and it is shown that in this case, unlike to
the case that only the coin is subjected to decoherence, the
entanglement converges to a significant value. Moreover, a closed
formula for the probability distribution has been presented in this
section. We summarize our results and present our conclusions in
section IV.

\section{Background}\label{background}
QW is the quantum version of the CRW where instead of coin flipping,
we use the coin operator to make superposition on the coin space,
and instead of the walking we use the translation operator to move
quantum particle according to the coin's degrees of freedom. In
one-dimensional quantum walk we have two degrees of freedom in the
coin space $H_c$, spanned by $\{|L\rangle,|R\rangle\}$, and infinite
degrees of freedom in the position space $H_p$, spanned by
$\{|i\rangle \,\,\,  i=-\infty,\cdots,\infty \}$. The walker Hilbert
space $H_W$ is defined as the tensor product of the coin space $H_c$
and the position space $H_p$, i.e. $H_W=H_s\otimes H_c$. Each step
of QW is constructed by the unitary operator $U_c$, making a
superposition on the coin space, followed by the translation
operator $S$, moving the particle according to the coin state, i.e
\begin{equation}\label{U-w}
U_w=S\;(I\otimes U_c),
\end{equation}
where
\begin{equation}\label{S}
S=\sum\limits_x {\left| {x + 1} \right\rangle \left\langle x \right|
\otimes \left| R \right\rangle \left\langle R \right| + \left| {x -
1} \right\rangle \left\langle x \right| \otimes \left| L
\right\rangle \left\langle L \right|}.
\end{equation}
Therefore, the unitary evolution of the quantum walker is as follows
\begin{equation}\label{t-step-waking}
|\Psi\left(t+1\right)\rangle={U_w}|\Psi\left(t\right)\rangle
\,\,\rightarrow
\,\,|\Psi\left(t\right)\rangle={U_w}^t|\Psi\left(0\right)\rangle.
\end{equation}
This is, in fact, the coherent evolution of the system but, in
practice, it is not possible to isolate the system from the
environment and, in general, the environment affects the coherent
evolution of the system. Generally, decoherence is used to estimate
deviation from the ideal case in which the effects of environment
are neglectable. One important approach to investigating decoherence
is the so-called Kraus representation \cite{Nielsen}. Let us define
$H_E$ as the Hilbert space of the environment, spanned by
$\{\ket{e_n}$, $n = 0\cdots m\}$ where $m$ is the dimension of the
environment's Hilbert space. Now, in order to study the time
evolution of the system we should consider the time evolution of the
whole system, i.e. system+environment, defined on the Hilbert space
$H=H_E\otimes H_W$, and obtain the state of the system by tracing
out over the environment's degrees of freedom, i.e.
\begin{equation}\label{rho-sys}
    \rho_{\sys}=\Tr_{\Env} \left( U \rho\, U^\dagger \right).
\end{equation}
Here $U$ acts both on the system and the environment Hilbert spaces.
Without loss of generality we assume that the state of the whole
system is $\rho=\rho_0\otimes|env_0\rangle\langle env_0|$. So we can
write Eq. (\ref{rho-sys}) as
\begin{equation}\label{KR-rho-rep-1}
\rho_{\sys}=\sum\limits_{n = 0}^m {\left\langle {e_n }
\right|U\left| {env_0 } \right\rangle \rho _0 \left\langle {env_0 }
\right|U^\dag \left| {e_n } \right\rangle }  = \sum\limits_{n = 0}^m
{E_n \rho_0 E_n^\dag  },
\end{equation}
where $E_n=\left\langle {e_n} \right|U\left| {env_0 } \right\rangle,
\;\; n=0,1,\cdots,m$ are the so-called Kraus operators. These
operators satisfy the following completeness relation
\begin{equation}\label{compeletness-relation}
\sum\limits_n{{E_n}^\dag E_n }=I.
\end{equation}
By definition of the Kraus operators, one step of the walking can be
written as follows
\begin{equation}\label{first-step-rho}
\rho\left(t+1\right)=\sum\limits_{n = 0}^m {E_n \rho\left(t\right)
E_n^\dagger}.
\end{equation}
For $t$ steps, we can write
\begin{equation}\label{KR-rho-rep-t}
\rho\left(t\right)=\sum\limits_{n_t = 0}^m {...\sum\limits_{n_2 =
0}^m {\sum\limits_{n_1 = 0}^m
{E_{n_t}...E_{n_2}E_{n_1}\rho\left(0\right) E_{n_1}^\dagger
E_{n_2}^\dagger ...E_{n_t}^\dagger}}}.
\end{equation}
It is worth noting that Eq. (\ref{KR-rho-rep-t}) is general and the
Kraus operators $E_n$ include the whole information of all types of
evolution. It follows, therefore, that the coin operator, the
translation operator and the environment effects all are embedded in
$E_n$ and we did not assume any restriction yet.

To make any progress, we should therefore find the Kraus operators
for our system defined in Eq. (\ref{KR-rho-rep-1}),  and use Eq.
(\ref{KR-rho-rep-t}) in order to obtain the final state $\rho(t)$.
Evidently,  the $E_n$  are operators that act on the system
(coin+position) Hilbert space, and therefore we can write the
general form of $E_n$ as follows
\begin{eqnarray}\label{general-En}
E_n  &=& \sum\limits_{x,x'}{\sum\limits_{i,j} {a_{x,x',i,j}^{(n)}
\left| {x'} \right\rangle \left\langle x \right| \otimes \left| i
\right\rangle \left\langle j \right|} }\\\nonumber &=& \sum\limits_x
{\sum\limits_l {\sum\limits_{i,j} {a_{x,l,i,j}^{(n)} \left| {x + l}
\right\rangle \left\langle x \right| \otimes \left| i \right\rangle
\left\langle j \right|} } },
\end{eqnarray}
where $x,l=-\infty,\cdots,\infty$ and $i,j=\{L,R\}$.

Recently, we have shown that a reasonable suggestion on the
coefficient $a^{(n)}$ of Eq. (\ref{general-En}) enables us to derive
useful analytical expression for the first and second moments of
position \cite{annabestani-deco}. In the remaining of this section,
we briefly introduce the method. To begin with, we use the Fourier
transformation
\begin{equation} \label{Four-x}
\left| x \right\rangle  = \int\limits_{ - \pi }^\pi
{\frac{{dk}}{{2\pi }}e^{ - ikx} \left| k \right\rangle },
\end{equation}
and write Eq. (\ref{general-En}) in the $k$-space as \small
\begin{equation}\label{En-Kspace}
\tilde{E_n}  = \sum\limits_{x,l} {\sum\limits_{i,j} {a_{x,l,i,j}^{(n)}
\iint {\frac{{dkdk'}}{{4\pi ^2 }}e^{ - ilk} e^{ - ix\left( {k - k'} \right)}
 \left| k \right\rangle \left\langle {k'} \right| \otimes \left| i \right\rangle \left\langle j \right|} }
 }.
\end{equation}
\normalsize If we assume that the coefficients $a^{(n)}_{x,l,i,j}$
are not dependent on the coordinate $x$, then one can use the
orthonormalization  relation $\sum\limits_x {e^{ - ix\left( {k - k'}
\right)} } = 2\pi \delta \left( {k - k'} \right)$ and caries out the
summation on $x$. This is, of course, a reasonable assumption  for a
large family of QW, including the QW on the homogeneous position
space. With this assumption, we have
\begin{equation}\label{En-kspace-finalform}
\tilde{E_n}  = \int {\frac{{dk}}{{2\pi }}\left| k \right\rangle
\left\langle k \right| \otimes C_n\left(k\right)},
\end{equation}
where
\begin{equation}\label{Cn}
C_n \left( k \right) = \sum\limits_l {\sum\limits_{i,j} {a_{l,i,j}^{(n)} e^{ - ilk} \left| i \right\rangle \left\langle j \right|}. }
\end{equation}
Now, if we write the general form of $\rho_0$ in the $k$-space as
\begin{equation} \rho _0  = \iint {\frac{{dkdk'}}{{4\pi ^2 }}\left|
k \right\rangle \left\langle {k'} \right| \otimes \left| {\psi _0 }
\right\rangle \left\langle {\psi _0 } \right|},
\end{equation}
we get, from Eq. (\ref{first-step-rho}),  the following form for the
first step of walking
\begin{eqnarray}\nonumber
\rho '  &=& \iint {\frac{{dkdk'}}{{4\pi ^2 }}\left| k \right\rangle \left\langle {k'} \right| \otimes \sum\limits_n {C_n\left(k\right)
 \left| {\psi _0 } \right\rangle \left\langle {\psi _0 } \right|C_n^\dag\left(k'\right)  } }  \\
&=&\iint {\frac{{dkdk'}}{{4\pi ^2 }}\left| k \right\rangle
\left\langle {k'} \right| \otimes \mathcal{L}_{k,k'}\left| {\psi _0
} \right\rangle \left\langle {\psi _0 } \right|  },
\end{eqnarray}
where the superoperator $\mathcal{L}_{k,k'}$ is introduced by
\begin{equation}\label{L-k,k'}
  \mathcal{L}_{k,k'} \tilde{O}= \sum\limits_n{C_n\left(k\right) {\tilde{O}} C_n^\dag\left(k'\right).}
\end{equation}
Therefore, after $t$ steps walking,  we find
\begin{equation}\label{rho-superop-t-step}
\rho\left(t\right)=\iint {\frac{{dkdk'}}{{4\pi ^2 }}\left| k
\right\rangle \left\langle {k'} \right| \otimes
\mathcal{L}^{t}_{k,k'}\left| {\psi _0 } \right\rangle \left\langle
{\psi _0 } \right|},
\end{equation}
as the state of the walker and
\begin{eqnarray}\label{probability(x,t)}
p\left( {x,t} \right) &=& \iint {\frac{{dkdk'}}{{4\pi ^2 }}
\scalar{x}{k}\scalar{k'}{x}\Tr\left( {\mathcal{L}_{kk'}^t \rho _0 }
\right)}\\ \nonumber &=& \iint {\frac{{dkdk'}}{{4\pi ^2 }}
e^{-ix\left(k'-k\right)}\Tr\left( {\mathcal{L}_{kk'}^t \rho _0 }
\right)},
\end{eqnarray}
as the probability of finding the walker in the position $x$. Note
that the completeness relation on the Kraus operators, given in Eq.
(\ref{compeletness-relation}), now implies that the coin operators
$C_n(k)$ satisfy the same relation, i.e.
\begin{equation}\label{completeness-Cn}
 \sum\limits_n{C^{\dag}_n(k) C_n(k)}=I,
\end{equation}
where  can be used to prove another important property of
$\mathcal{L}_{k,k}$, i.e.  the \textit{trace preserving} condition
\begin{equation}\label{trace-preserving}
\Tr\left( \mathcal{L}^{m}_{k,k} \tilde O \right) =  \Tr\left(
{\tilde O} \right),
\end{equation}
for arbitrary operator $\tilde{O}$.

By definition, the $m$th moment of the probability distribution
$p\left(x,t\right)$ is as follows
\begin{equation}\label{m-moment}
\left\langle {x^m } \right\rangle  = \sum\limits_x {x^m p\left(
{x,t} \right)},
\end{equation}
where can be used , together with Eq. (\ref{probability(x,t)}),  and
write the first and second moments  as
\begin{eqnarray}\label{moments-1,2}
\left\langle {x } \right\rangle  &=& \frac{i}{{2\pi  }}\iint {dkdk'
\frac{d\delta}{dk} \Tr\left( {\mathcal{L}_{kk'}^t \rho _0 } \right)}
\\ \nonumber \left\langle {x^2 } \right\rangle  &=& \frac{1}{{2\pi
}}\iint {dkdk' \frac{d^2\delta}{dk'dk} \Tr\left(
{\mathcal{L}_{kk'}^t \rho _0 } \right)}.
\end{eqnarray}
Now, by using the relations
\begin{eqnarray}\label{diff-L-G-rep}
\frac{d}{{dk}}\Tr\left( {\mathcal{L}_{kk'} \rho} \right) = \Tr\left(
{\mathcal{G}_{kk'} \rho} \right) \\\nonumber
\frac{d}{{dk'}}\Tr\left( {\mathcal{L}_{kk'} \rho} \right) =\Tr\left(
{\mathcal{G^\dag}_{k' k} \rho} \right),
\end{eqnarray}
where
\begin{equation}\label{G}
\mathcal{G}_{k,k'}\tilde{O}=\sum\limits_n {\frac{{dC_n \left( k
\right)}}{{dk}}\tilde{O}C^{\dag}_n \left(k'\right)},
\end{equation}
one can carries out the integrations of Eq. (\ref{moments-1,2}) and,
after some calculations, get the following relations for  the first
and second moments \small
\begin{flushleft}
\begin{eqnarray}\label{moments-1,2-final-form}
{\langle x\rangle} _t &=&  i \int_{ - \pi }^\pi \frac{dk}{{2\pi}}
\sum\limits_{m = 1}^t \Tr\left\{ {{\cal G}_k \left( {{\cal L}_k^{m -
1} \left| {\psi _0 \rangle \langle \psi _0 } \right|} \right)}
\right\} \\ \nonumber \left\langle {x^2 } \right\rangle_t  &=&
\int_{ - \pi }^\pi  \frac{dk}{2\pi}\sum\limits_{m = 1}^t
\sum\limits_{m' = 1}^{m - 1} \Tr\Big\{ \mathcal{G}_k^\dag
\mathcal{L}_k^{m - m' - 1} \left( \mathcal{G}_k \mathcal{L}_k^{m' -
1} \left|\psi_0\rangle\langle\psi_0\right|  \right)\\\nonumber &&+
\mathcal{G}_k \mathcal{L}_k^{m - m' - 1} \left( \mathcal{G}_k^\dag
\mathcal{L}_k^{m' - 1} \left|\psi_0\rangle\langle\psi_0\right|
\right)\Big\}  \\\nonumber &&+ \int_{ - \pi }^\pi
{\frac{dk}{{2\pi}}\sum\limits_{m = 1}^t {\Tr\left\{
{\mathcal{J}_{k}\left( {\mathcal{L}_{k}^{m - 1}
\left|\psi_0\rangle\langle\psi_0\right| } \right)}  \right\}}}.
\end{eqnarray}
\end{flushleft} \normalsize
Here we have defined, for simplicity,
$\mathcal{L}_{k}\equiv\mathcal{L}_{k,k}$,
$\mathcal{G}_{k}\equiv\mathcal{G}_{k,k}$ and
\begin{equation}\label{J}
\mathcal{J}_{k}=\left.\frac{{d\mathcal{G}_{k,k'}^\dag
}}{{dk}}\right|_{k'=k}=\left.\sum\limits_n {\frac{{dC_n \left( k
\right)}}{{dk}}\tilde{O}\frac{{dC^{\dag}_n \left( k'
\right)}}{{dk'}}} \right|_{k'=k}.
\end{equation}

\section{Calculations and results}
In this section, by using the analytic expressions for the first and
second moments given in Eq. (\ref{moments-1,2-final-form}), we
investigate the \textit{variance} of the 1DQW in the presence of
tunneling noise, i.e. a type of decoherence which is applied on the
position subspace.

\subsection{Variance of decoherent 1DQW}
Consider a one-dimensional QW such that after each step of walking, the walker can move to the
nearest neighbors with probability \textit{p}. This phenomena occurs
in the experimental setup because of tunneling effect. The effect of
such noise on the probability distribution has investigated,
numerically, by D\"{u}r \textit{et al} \cite{Dur-prop}. In this
section we calculate, analytically, the variance of 1DQW in the
presence of this kind of noise and investigate some important
features of it.

Let us assume that, after each step in 1DQW, the walker tunnels to
the left or to the right site with probability \textit{p}, so it is
clear that the walker does not move with probability $(1-p)$.
Therefore, unlike the coherent QW, all sites may be occupied. In the
presence of such tunneling, one step of the walking can be written
as
\begin{eqnarray}\label{one_step_decoherent_rho}
\rho\left( t+1\right)  &=& \left(1-p\right)U_w \rho\left(
t\right)U^{\dag}_w \\ \nonumber &+&\frac{p}{2}\left(\mathcal{S}_+
U_w \rho\left(t\right) U^{\dag}_w \mathcal{S}^{\dag}_+ +
\mathcal{S}_- U_w  \rho\left(t\right) U^{\dag}_w
\mathcal{S}^{\dag}_- \right),
\end{eqnarray}
where $ \mathcal{S}_\pm =\sum\limits_x {\left| {x \pm 1 }
\right\rangle \left\langle {x } \right| \otimes I_c }$ and $U_w$ is
the coherent walking operator, given by Eq. (\ref{U-w}). For $p=0$,
the system is exactly the same as the coherent QW and the evolution
of the system is unitary, so the final state of the system remains
pure. On the other hand, for $p \ne 0$ the evolution is nonunitary
and decoherenc is happened.

\begin{figure}
\centering
\includegraphics[width=9 cm]{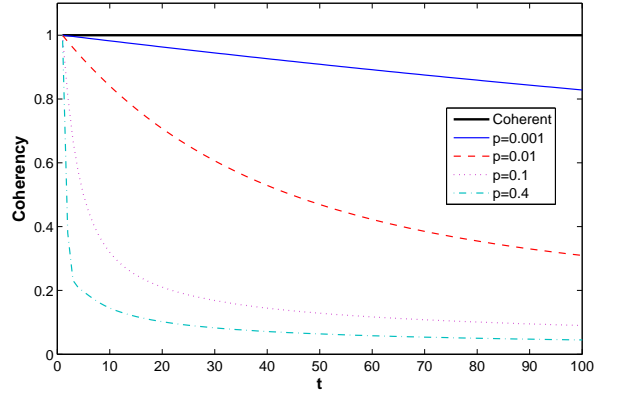}
\caption{(Color online) Coherency of system versus \textit{t} for
various $p$.  Note that for $p=0$ (coherent case), the coherency
does not decay.} \label{Fig:coherency}
\end{figure}

Generally, decoherence can be considered as a symbol to express the
reduction of purity of the system state and, therefore, one can use
the linear entropy $S(\rho(t))=1-\Tr[\rho^2(t)]$ as a measure of
decoherence. The linear entropy has the limiting values 0 and $1 -
1/N$, respectively, for pure and maximally mixed states, where $N$
is the dimension of the space that the density matrix $\rho(t)$ is
supported on. In the same manner, one can use the purity (or
coherency)
\begin{equation}
    \mathcal{C}\left(t\right)=\Tr\left[\rho^2\left(t\right)\right],
\end{equation}
where has the limiting values 1 and $1/N$ for pure and maximally
mixed states, respectively. Therefore, the purity less than 1 is a
signature of decoherent evolution.   In Fig.\ref{Fig:coherency}, we
have plotted the coherency of the system as a function of time in
the presence of the tunneling noise of Eq.
(\ref{one_step_decoherent_rho}), for different values of the
probability $p$.  As we expect,  for $p=0$ the evolution of QW takes
place coherently, while by increasing  $p$ the system loses its
coherency very fast.

Now by comparing  Eq. (\ref{one_step_decoherent_rho}) with Eq.
(\ref{first-step-rho}), we can write the Kraus operators as
\begin{eqnarray}\label{Kraus_operators}
    E_1=\sqrt{1-p}U_w\\
    E_2=\sqrt{\frac{p}{2}}\mathcal{S}_+ U_w\\
    E_3=\sqrt{\frac{p}{2}}\mathcal{S}_- U_w.
\end{eqnarray}
All the above Kraus operators have the form given by Eq.
(\ref{general-En}), therefore, by using Eq. (\ref{Cn}) we  find the
$C_i$ matrices as
\begin{eqnarray}\label{C-position}
C_1&=&\sqrt {\frac{{1 - p}}{2}} \left( {\begin{array}{*{20}c}
   {e^{ - ik} } & {e^{ - ik} }  \\
   {e^{  ik}} & {-e^{  ik}}  \\
\end{array}} \right)\\\nonumber
C_2&=&\frac{\sqrt {p}}{2} \left( {\begin{array}{*{20}c}
   {e^{ - 2ik} } & {e^{ - 2ik} }  \\
   {1 } & {-1}  \\
\end{array}} \right)\\\nonumber
C_3&=&\frac{\sqrt {p}}{2}\left( {\begin{array}{*{20}c}
    {1} & {  1}  \\
    {e^{ 2ik} } &  {-e^{ 2ik} }  \\
\end{array}} \right),
\end{eqnarray}
where satisfy, clearly,  Eq. (\ref{completeness-Cn}).

Now by using the affine map \cite{Nielsen}, we can represent
arbitrary two-by-two matrix $\tilde{O}$ by a four-dimensional column
vector as
\begin{equation}\label{affine-map}
     \tilde{O}=\sum\limits_{i=0}^3{r_i\sigma_i},
\end{equation}
where $r_i=\Tr(\sigma_i \tilde{O})/2$. Now, in order to find the
superoperators $\mathcal{L}_k , \mathcal{G}_k ,
\mathcal{G}^{\dag}_k$ and $ \mathcal{J}_k$, we use the affine map
and represent the action of them on an arbitrary two-by-two matrix
$\tilde{O}$ as follows \small
\begin{equation}\label{L-position}
\mathcal{L}_k\tilde{O}=\left( \begin {array}{*{20}c} 1&0&0&0\\
0&0& \sin \left(2k \right) & \cos \left( 2k \right)\\
0&0& -\cos \left( 2k \right)& \sin \left( 2k\right) \\
0&1&0&0  \end {array} \right) \left( {\begin{array}{*{20}c}
   {r_0 }  \\
   {r_1 }  \\
   {r_2 }  \\
   {r_3 }  \\
\end{array}} \right)
\end{equation}
\begin{equation}\label{G-position}
\mathcal{G}_k\tilde{O}=\left( \begin {array}{*{20}c} 0&-i&0&0\\
0&0& \cos \left(2k \right) & -\sin \left( 2k \right)\\
0&0& \sin \left( 2k \right)& \cos \left( 2k\right) \\
-i&0&0&0  \end {array} \right) \left( {\begin{array}{*{20}c}
   {r_0 }  \\
   {r_1 }  \\
   {r_2 }  \\
   {r_3 }  \\
\end{array}} \right)
\end{equation}
\begin{equation}\label{J-position}
\mathcal{J}_k\tilde{O}=\left( \begin {array}{*{20}c} 1+p&0&0&0\\
0&0& q\sin \left(2k \right) & q\cos \left( 2k \right)\\
0&0& -q\cos \left( 2k \right)& q\sin \left( 2k\right) \\
0&1+p&0&0  \end {array} \right) \left( {\begin{array}{*{20}c}
   {r_0 }  \\
   {r_1 }  \\
   {r_2 }  \\
   {r_3 }  \\
\end{array}} \right),
\end{equation}
and $\mathcal{G}_k^\dag=\mathcal{G}^\ast_k$, which is obtained from
the Hermiticity of the Pauli matrices.
\normalsize

Now, according to   Eq. (\ref{moments-1,2-final-form}), in order to
calculate moments and variance, we need to find the $m$th power of
$\mathcal{L}_k$. But, here, the method  of non-trivial submatrix
$M_k$, which is introduced in \cite{Brun03}, is not applicable
because $\det\left(I-M\right)=0$, i.e. $I-M$ is not invertible.
Fortunately, the eigenvalues and eigenvectors of $\mathcal{L}_k$ are
enough simple to calculate $\mathcal{L}^m_k$ directly.
Straightforward calculations show that the eigenvalues and the
corresponding eigenvectors of matrix $\mathcal{L}_k$, given in Eq.
(\ref{L-position}), are as follows, respectively
\begin{equation}\label{eigenvalues-position}
\lambda_1=\lambda_2=1
\,\,\,\,,\,\,\,\,\,\,\lambda_3=e^{i\left(\theta+\pi\right)}\,\,\,\,,\,\,\,\,\,\,\,\lambda_4=e^{-i\left(\theta+\pi\right)},
\end{equation}
and  \small
\begin{equation}\label{eigenvectors-position}
\begin{array}{l}
\left|e_1\right\rangle=\frac{1}{N_1}\left( \begin {array}{c} \sin \left( k \right) \\
\cos \left( k \right) \\\sin \left( k \right)
\\\cos \left( k \right) \end {array} \right) \\ \left|e_2\right\rangle=\frac{1}{N_2}\left( \begin {array}{c} -2\left( \cos^2\left(k\right)+1 \right) \\
\sin \left( 2k \right) \\2\sin^2 \left( k \right)
\\\sin \left( 2k \right) \end {array} \right) \\
\left|e_3\right\rangle=\left|e_4\right\rangle^*=\frac{1}{N_3}\left( \begin {array}{c} 0 \\
\sin \left( 2k \right)e^{2i\theta} \\2\sin^2 \left( k \right)e^{2i\theta}
\\-\sin \left( 2k \right)e^{i\theta} \end {array} \right),
\end{array}
 \end{equation}
\normalsize where  $\cos\left(\theta\right)=\cos^2\left(k\right)$
and $N_i$ are the normalization factors. By the help of eigenvectors
and the spectral decomposition, we are able to find any power of
matrix $\mathcal{L}_k$, appeared in Eq.
(\ref{moments-1,2-final-form}).

In order to calculate the first moment  $\langle x \rangle$,  we
need \small
\begin{eqnarray}\label{sum-L-in-moment1}
\nonumber
\Gamma&=&\sum\limits^t_{m=1}{\mathcal{L}^{m-1}_k}=\sum\limits^t_{m=1}
\Big\{\left|e_1\right\rangle\left\langle
e_1\right|+\left|e_2\right\rangle\left\langle e_2\right|\\\nonumber
&&+e^{i\left(\theta+\pi\right)\left(m-1\right)}\left|e_3\right\rangle\left\langle
e_3\right|+e^{-i\left(\theta+\pi\right)\left(m-1\right)}\left|e_4\right\rangle\left\langle
e_4\right|\Big\}\\\nonumber
&=&t\left(\left|e_1\right\rangle\left\langle
e_1\right|+\left|e_2\right\rangle\left\langle
e_2\right|\right)\\\nonumber
&&+\left(\frac{1-e^{it\left(\theta+\pi\right)}}{1+e^{i\theta}}\right)
\left|e_3\right\rangle\left\langle e_3\right|+
\left(\frac{1-e^{-it\left(\theta+\pi\right)}}{1+e^{-i\theta}}\right)\left|e_4\right\rangle\left\langle e_4\right|\\
&=&t\left(\left|e_1\right\rangle\left\langle e_1\right|
+\left|e_2\right\rangle\left\langle e_2\right|\right)+2\Re{\left(\frac{1}{1+e^{i\theta}}\left|e_3\right\rangle\left\langle e_3\right|\right).}
\end{eqnarray}
\normalsize Here, in the last equality,  we used the fact that the
forth eigenvalue and eigenvector are complex conjugate of the third
one, respectively (see Eqs. (\ref{eigenvalues-position}) and
(\ref{eigenvectors-position})). Also we neglect the oscillatory term
$e^{it\left(\theta+\pi\right)}$, because according to the phase
stationary theorem  the oscillatory term can be neglected in long
time limit.

By putting  Eq. (\ref{eigenvectors-position}) into Eq.
(\ref{sum-L-in-moment1}) and some simplification, we find the
following matrix form for $\Gamma$
\begin{widetext}
\begin{equation}
\Gamma=\sum\limits^{t}_{m=1}{\mathcal{L}^{m-1}_k}=\frac{1}{\Delta}
\left( \begin {array}{*{20}c} 2t\left[ \cos^2\left(k\right)+1\right]&0&0&0\\
0& 2t\cos^2\left(k\right)+1 & t\sin \left(2k \right) &  2t\cos^2\left(k\right)-1 \\
0& \left(t-1\right)\sin \left(2k \right) & \cos^2 \left( k \right)+t\sin^2 \left( k \right)& t\sin \left( 2k\right) \\
0&2\left(t-1\right)\cos^2\left(k\right)+1& \left(t-1\right)\sin
\left(2k \right) & 2t\cos^2\left(k\right)+1  \end {array} \right),
\end{equation}
\end{widetext}
where $\Delta=2\left[ \cos^2\left(k\right)+1\right]$. We can,
therefore, write the first moment $\langle x \rangle$, given  in Eq.
(\ref{moments-1,2-final-form}), as follows
\begin{eqnarray}\label{moments-1-final-result-position}
\langle x\rangle _t &=&  {\frac {i}{ 2\pi}}
\int_{ - \pi }^\pi  {dk \left( {\begin{array}{*{20}c}
   {0} & {-2i} & {0} & {0}  \\
\end{array}} \right)\Gamma\left( {\begin{array}{*{20}c}
   {r_0}  \\
   {r_1}  \\
   {r_2 }  \\
   {r_3 }  \\
\end{array}} \right)}\\\nonumber
&=& \left[\left(2-\sqrt {2}\right)t+\frac{1}{\sqrt {2}} \right]
r_1+\left[\left(2-\sqrt {2}\right)t-\frac{1}{\sqrt {2}} \right] r_3.
\end{eqnarray}
Note, however,  that in calculating  this   moment from Eq.
(\ref{moments-1,2-final-form}) only the first row of $\mathcal{G}_k$
gives the nonzero contribution. Clearly, the above expression for
$\langle x\rangle _t$ does not contain $p$, which implies that this
type of noise can not change the first moment.

Now, in order to calculate $\langle x^2 \rangle$, we first calculate
the last term of Eq. (\ref{moments-1,2-final-form}) as follows
\begin{eqnarray}\label{last-term-moment2-position}
\nonumber &&\frac{1}{{2\pi }} \int_{ - \pi }^\pi dk{\sum\limits_{m =
1}^t {\Tr\left\{  {\mathcal{J}_{k}\left( {\mathcal{L}_{k}^{m - 1}
\left|\psi_0\rangle\langle\psi_0\right| } \right)}
\right\}}}\\\nonumber &=&{\frac {1}{ \pi}} \int_{ - \pi }^\pi  {dk
\left( {\begin{array}{*{20}c}
   {1+p} & {0} & {0} & {0}  \\
\end{array}} \right)\Gamma\left( {\begin{array}{*{20}c}
   {r_0}  \\
   {r_1}  \\
   {r_2 }  \\
   {r_3 }  \\
\end{array}} \right)}\\
&=&t\left(1+p\right).
\end{eqnarray}
On the other hand, the first term of Eq.
(\ref{moments-1,2-final-form}) can be written as follows
\begin{widetext}
\begin{eqnarray}\label{first-term-moment2-position}
\int_{ - \pi }^\pi  {\frac{dk}{{2\pi}}\sum\limits_{m = 1}^t
{\sum\limits_{m' = 1}^{m - 1} {\Tr\left\{ {\mathcal{G}_k^\dag
\mathcal{L}_k^{m - m' - 1} \left( {\mathcal{G}_k \mathcal{L}_k^{m' -
1} \left|\psi_0\rangle\langle\psi_0\right| } \right)} +
\mathcal{G}_k \mathcal{L}_k^{m - m' - 1} \left( {\mathcal{G}_k^\dag
\mathcal{L}_k^{m' - 1} \left|\psi_0\rangle\langle\psi_0\right| }
\right)\right\}}}}\\\nonumber =\frac{1}{{\pi }}\int_{ - \pi }^\pi
{dk\sum\limits_{m = 1}^t {\sum\limits_{m' = 1}^{m - 1}
\left({\begin{array}{*{20}c}
 {0} & { i } & 0 & 0  \\
\end{array}}\right) {\mathcal{L}_k^{m - m' - 1}
\left( \mathcal{G}_k -\mathcal{G}_k^{\dag}\right) \mathcal{L}_k^{m' - 1} \left( {\begin{array}{*{20}c}
   {1/2 }  \\
   {r_1 }  \\
   {r_2 }  \\
   {r_3 }  \\
\end{array}} \right)  } }},
\end{eqnarray}
\end{widetext}
where we have used the fact that
$\mathcal{G}_k^{\dag}=\mathcal{G}_k^{*}$. From Eq.
(\ref{G-position}) the exact form of
$\mathcal{G}_{k}-\mathcal{G}^{\dag}_{k}$ is
\begin{equation}\label{G-Gdag}
\left(\mathcal{G}_k-\mathcal{G}^{\dag}_k\right)\tilde{O}=\left( \begin {array}{*{20}c} 0&-2i&0&0\\
0&0&0&0\\
0&0&0&0 \\
-2i&0&0&0  \end {array} \right) \left( {\begin{array}{*{20}c}
   {r_0 }  \\
   {r_1 }  \\
   {r_2 }  \\
   {r_3 }  \\
\end{array}} \right).
\end{equation}
Also from the trace preserving property of $\mathcal{L}_k$, we have
\begin{equation}\label{Lm-tracepreserv}
\mathcal{L}_k^{m-1}\left( {\begin{array}{*{20}c}
   {1/2 }  \\
   {r_1 }  \\
   {r_2 }  \\
   {r_3 }  \\
\end{array}} \right)=\left( {\begin{array}{*{20}c}
   {1/2 }  \\
   {r'_1 }  \\
   {r'_2 }  \\
   {r'_3 }  \\
\end{array}} \right).
\end{equation}
So, easily, we get
\begin{equation}
\left( \mathcal{G}_{k}-\mathcal{G}^{\dag}_{k}\right)\mathcal{L}_k^{m'-1}\left|\psi_0\rangle\langle \psi_0\right|=\left( {\begin{array}{*{20}c}
   {-2ir'_1 }  \\
   {0 }  \\
   {0 }  \\
   {-i}  \\
\end{array}} \right).
\end{equation}
Putting this into Eq. (\ref{first-term-moment2-position}), we get
\begin{equation}\label{moments2-lastterm-befor-final-integration}
\int_{ - \pi }^\pi  {\frac{dk}{\pi} \left({\begin{array}{*{20}c}
 {0} & { 1 } & 0 & 0  \\
\end{array}}\right) \Gamma ' \left( {\begin{array}{*{20}c}
   {0 }  \\
   {0 }  \\
   {0 }  \\
   {1 }  \\
\end{array}} \right)},
\end{equation}
where $\Gamma'$ is defined by
\begin{eqnarray}
\nonumber &&\Gamma '=\sum\limits_{m = 1}^t {\sum\limits_{m' = 1}^{m
- 1}\mathcal{L}_k^{m - m' - 1}}=\sum\limits_{m = 1}^t
\sum\limits_{m' = 1}^{m - 1}
\Big\{\left|e_1\right\rangle\left\langle
e_1\right|+\left|e_2\right\rangle\left\langle e_2\right|\\\nonumber
&&+e^{i\left(\theta+\pi\right)\left(m-m'-1\right)}\left|e_3\right\rangle\left\langle e_3\right|
+e^{-i\left(\theta+\pi\right)\left(m-m'-1\right)}\left|e_4\right\rangle\left\langle e_4\right|\Big\}\\
&&=\frac{t}{2}\left(t-1\right)\left\{\left|e_1\right\rangle\left\langle e_1\right|+\left|e_2\right\rangle\left\langle e_2\right|\right\}\\
&&+2\Re{\left\{
\left(\frac{t}{1+e^{i\theta}}-\frac{1}{\left(1+e^{i\theta}\right)^2}\right)\left|e_3\right\rangle\left\langle
e_3\right|\right\}}.
\end{eqnarray}
From Eq. (\ref{moments2-lastterm-befor-final-integration}) it  is
clear that we need only the term $\Gamma_{2,4}'$, which is
\begin{equation}
\Gamma_{2,4}'=\frac {t^2\left(\cos^4 \left( k \right)+ \cos^2 \left(
k\right)\right)+1}{ 2\left( \cos^2 \left( k \right)+1 \right)
^2}-\frac{t}{2}.
\end{equation}
Using this into Eq.
(\ref{moments2-lastterm-befor-final-integration}), caring the
integration and putting everything together, we find the following
form for the second moment
\begin{eqnarray}\label{moments2-position-final-resualt}
\nonumber
\langle x^2 \rangle &=& t\left(1+p\right)+\left( 1-\frac{1}{\sqrt {2}}\right) t^2-t+\frac{3\sqrt {2}}{8}\\
&=&\left( 1-\frac{1}{\sqrt {2}}\right) t^2+ tp+\frac{3\sqrt {2}}{8}.
\end{eqnarray}
It is worth to note that the second moment does not depend on the
initial state and, surprisingly, it involves the quadratic term
$t^2$. This means that the decoherent 1DQW preserves the quantum
property (ballistic behavior of variance) whenever the position
space is subjected to decoherence (tunneling noise), in contrary to
the case that the coin is subjected to noise in which for long time
walking the system becomes  classic and its variance becomes linear,
even for weak strength of noise \cite{Brun03}.

\begin{figure}
\centering
\includegraphics[width=9 cm]{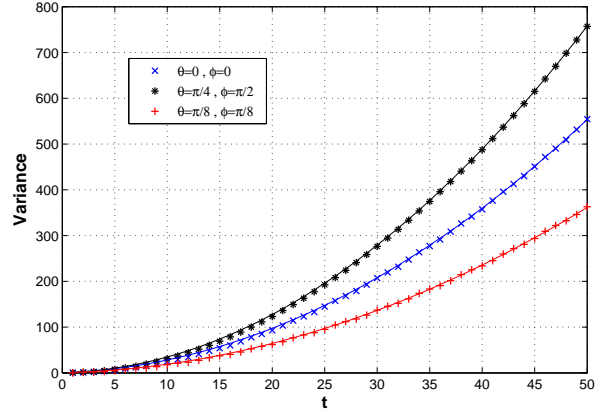}
\caption{(Color online) Variance versus \textit{t} for different
initial states of Eq. (\ref{general-initialState}). The continues
curves are the results of theoretical formula (\ref{variance}) and
the corresponding discrete symbols are the direct numerical
calculations. } \label{Fig:Variance}
\end{figure}

Let us look at the variance more precisely. From Eqs.
(\ref{moments-1-final-result-position}) and
(\ref{moments2-position-final-resualt}) we have
\begin{equation}\label{variance}
    V=\langle x^2 \rangle -\langle x \rangle ^2=A t^2+ B t+C,
\end{equation}
where
\begin{eqnarray}\label{varianceCoefficients}\nonumber
    A&=&\alpha-4\alpha^{2} \left( r_3+r_1 \right) ^2 \\\nonumber
    B&=& 2\sqrt{2}\alpha \left(r_3^2-r_1^2 \right) +p \\
    C&=& -\left( r_3-r_1 \right) ^2/2+3\sqrt {2}/8,
\end{eqnarray}
in which $\alpha=1-1/\sqrt{2}$. In the long time limit and for $A
\ne 0$, we can neglect the $B$ and $C$ coefficients. The question
that may arise here is that: \textit{is it possible to find initial
states for which $A=0$ and $B \ne 0$?} or, in other words:
\textit{is it possible to find initial states such that the variance
is classic?}. It is not difficult to show that it is impossible to
have $A=0$ even for mixed initial states. Let us consider the
following  general form for the initial state $\rho_0$
\begin{equation}
    \rho_0=\sum\limits_{i=0}^{3}{\sigma_ir_i}=\left( \begin{array}{cc} r_0+r_3&r_1-ir_2 \\ {r_1+ir_2}&r_0-r_3
    \end{array}\right),
\end{equation}
where $r_0=1/2$, because of the normalization, and the positivity of
$\rho_0$ implies that
\begin{equation}
    r_1^2+r_2^2+r_3^2 \leq \frac{1}{4}\; \Longrightarrow \; r_1^2+r_3^2 \leq
    \frac{1}{4}.
\end{equation}
This means that $|r_1|,|r_3| \leq 1/2$ or $2r_1 r_3\le 1/2$, so
\begin{equation}\label{positivity-condition}
    \left(r_1+r_3\right)^2=r_1^2+r_3^2+2r_1r_3 \leq \frac{3}{4}.
\end{equation}
On the other hand, the coefficient $A$ in Eq.
(\ref{varianceCoefficients}) vanishes only for
\begin{equation}
    \left(r_1+r_3\right)^2=\frac{1}{4\alpha} \approx 0.85,
\end{equation}
which is, obviously, in conflict with the condition given by Eq.
(\ref{positivity-condition}). Consequently, the tunneling noise
which is a noise on the position subspace of QW, not only dose not
change the quadratic behavior of variance ($p$ does not appear in
the coefficient $A$ in Eq. (\ref{varianceCoefficients})), but also
increases it a bit (as a linear term in the coefficient $B$).
Furthermore, although the quadratic term depends on the initial
state, but it can never be vanished by tuning the initial state and
it ranges between the minimum $\alpha -3\alpha^2$ and the maximum
$\alpha$. Now, in order to find the initial state for which the
variance is maximum,  let us consider the following generic form for
the pure initial state
\begin{equation}\label{general-initialState}
    |\psi_0\rangle=\cos{\theta}|R\rangle+e^{i\phi}\sin{\theta}|L\rangle,
\end{equation}
where $0 \leq \theta \leq \pi$ and $0 \leq \phi < 2\pi$. For this
initial state, the elements of $r_i$ in affine map representation
defined in Eq. (\ref{affine-map}) are as follows
\begin{eqnarray}
    \begin{array}{ll}
        r_0=\frac{1}{2} & r_1=\frac{1}{2}\cos{\phi}\sin{2\theta}\\
        r_2=\frac{1}{2}\sin{\phi}\sin{2\theta} & r_3=\frac{1}{2}\cos{2\theta}.
    \end{array}
\end{eqnarray}
Using these $r_1,r_3$ in the definition of $A$, given in Eq.
(\ref{varianceCoefficients}), and maximizing it, we find that for
all $\theta, \phi$, satisfying
$\phi=-\cos^{-1}\left(\cot{2\theta}\right)$, the variance is
maximum. In Fig. \ref{Fig:Variance}, we have plotted variance versus
time for various initial states and have compared the results with
the numerical simulation. The figure shows that there are excellent
matching between the theoretical result of Eq. (\ref{variance}) and
the direct numerical calculations. An interesting point which we
would like emphasize here is that, although our expression have
derived for long time limit, but very fast decay of oscillatory term
implies that we have very good  matching even after a few steps (see
Fig. (\ref{Fig:Variance})).

\subsection{Entanglement in decoherent QW}
In the previous subsection we showed that the tunneling noise
preserves the quantum property of variance. In this subsection we
turn our attention to another quantum property of the system,
namely, the entanglement between the coin and the position degrees
of freedom, in the presence of tunneling noise. We also compare the
result by the case that only the coin is subjected to decoherence.
The model that we choose for the coin decoherence is defined by
\begin{eqnarray}\label{one_step_CoinDecoherent_rho}
\rho\left( t+1\right)  &=& \left(1-p\right)U_w \rho\left(
t\right)U^{\dag}_w\\\nonumber &+&p\left[P_R U_w \rho\left(t\right)
U^{\dag}_w P_R + P_L U_w  \rho\left(t\right) U^{\dag}_w P_L \right],
\end{eqnarray}
where $P_R=I_x \otimes |R\rangle\langle R|$ and $P_L=I_x \otimes
|L\rangle\langle L|$ are projections on the coin subspace. Equation (\ref{one_step_CoinDecoherent_rho}) describes the situation in
which, after each step of walking a measurement can perform on the
coin subspace of the walker with probability $p$. The probability
distribution and the variance of this type of decoherence have been
investigated in Ref. \cite{Brun03}.

As we mentioned  before, in the presence of noise,  the state of the
system becomes mixed and, therefore, we use the negativity of the
partial transpose of the density matrix as a suitable measure of
entanglement which is, of course, a computable measure for mixed
bipartite systems \cite{Vidal_Negativity}. Negativity is based on
the Peres-Horodecki criterion for separability
\cite{peres,horodecki} and measures the degree to which the partial
transpose of $\rho$ fails to be positive, i.e. the absolute value of
the sum of negative eigenvalues of the partial transpose of the
density matrix
\begin{equation}
    N=\frac{1}{2}\left(\sum\limits_{i}{|\lambda_i '|}-1\right)=\sum\limits_{i}{|{\lambda^{'}_i}^{-}|},
\end{equation}
where $\lambda_i '$ and ${\lambda^{'}_i}^{-}$ are eigenvalues and
negative eigenvalues of the matrix $\rho '(t)$. Here $\rho '(t)$ is
the partial transpose of $\rho(t)$, which is obtained from the
density matrix $\rho\left(t\right)$ by taking the transpose of
$\rho(t)$ with respect to one of the subsystems, say the first
subsystem, i.e.
\begin{equation}
    \rho_{xc,yb}'=\rho_{xb,yc},
\end{equation}
where $x, y$ and $b, c$ denote  the position and the coin state
indices, respectively.

\begin{figure}
\centering
\includegraphics[width=9 cm]{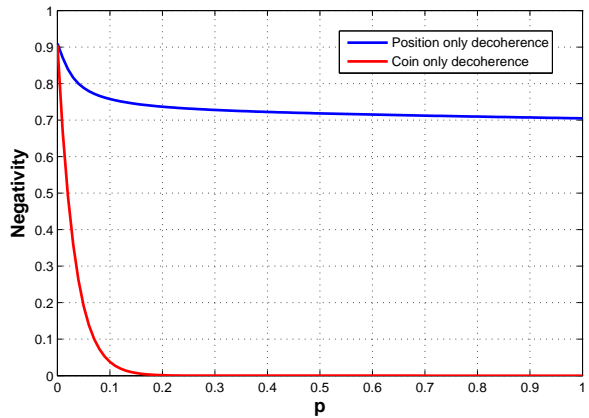}
\caption{(Color online) Negativity of the system with respect to
\textit{t} in the presence of the position only decoherence Eq.
(\ref{one_step_decoherent_rho}) (thick curve) and the coin only
decoherence Eq. (\ref{one_step_CoinDecoherent_rho}) (thin curve).}
\label{Fig:NegativityVsp}
\end{figure}

\begin{figure}
\centering
\includegraphics[width=9.5 cm]{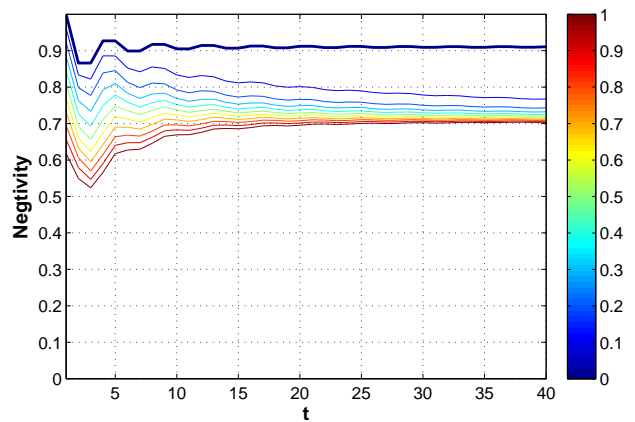}
\caption{(Color online) Negativity of the system versus $t$, in the
presence of tunneling noise.
 The curves from top to bottom show the negativity when the noise strength $p$ goes from $0$ to $1$ with $0.1$ step length.
 The upper curve (thick curve) is the negativity of the coherent case, i.e. $p=0$}
\label{Fig:NegativityVst}
\end{figure}

In Fig. (\ref{Fig:NegativityVsp}), we have plotted negativity of the
system in terms of the noise strength $p$, both for the cases that
the position is subjected to the noise given in Eq.
(\ref{one_step_decoherent_rho}) and the coin is subjected to the
noise given in Eq. (\ref{one_step_CoinDecoherent_rho}). Comparing
two negativities, it is clear from the figure that the system with
coin only decoherence is more influenced by the noise and losses its
negativity very fast. Moreover, though the negativity of the system
with coin only decoherence decays to zero value as $p$ grows up, but
situation is very different for the  system with position only
decoherence and, indeed, for such system the negativity converge to
a significant value, never less than $\approx 0.7$, even for maximum
rate of noise.
This means that the system with tunneling noise never loses all
quantum behavior, but  the system with  coin decoherence will do. In
Fig. (\ref{Fig:NegativityVst}), we have plotted negativity of the
system  as a function of time, in the presence of tunneling noise
for various $p$. The figure, clearly, shows the convergence behavior
of the negativity for this type of noise.
Note that the flat behavior of the negativity in coherent case, i.e.
$p=0$, changes to descending (ascending) behavior for $p<0.5$
($p>0.5$) (see Fig. \ref{Fig:NegativityVst}).

\subsection{Probability distribution of decoherent QW}

In this subsection we derive an analytical expression for the
probability distribution of decoherent quantum walk in terms of the
corresponding coherent probability distributions. To begin with, we
first rewrite Eq. (\ref{probability(x,t)}) in the coherent case
$p=0$ as
\begin{equation}\label{probability_0}
P_0\left(x,t \right) = \iint {\frac{{dkdk'}}{{4\pi ^2 }}
e^{-ix\left(k'-k\right)}\Tr\left( {\mathcal{W}_{kk'}^t \rho _0 }
\right)}.
\end{equation}
Note that we use $P_0\left(x,t \right)$ and $\mathcal{W}_{k,k'}$
instead of  $P\left(x,t \right)$ and $\mathcal{L}_{k,k'}$,
respectively,  in order to denote the corresponding quantities for
coherent QW. It is clear from Eqs. (\ref{L-k,k'}) and
(\ref{C-position}) that
\begin{equation}
    \mathcal{W}_{k,k'}\rho_0=U\left(k\right)\rho_0
    U^{\dag}\left(k'\right),
\end{equation}
where $U\left(k\right)$ is the Fourier form of the unitary
transformation of the Hadamard walk in the $k$-space, i.e.
\begin{equation}
    U\left(k\right)=\frac{1}{\sqrt{2}} \left( {\begin{array}{*{20}c}
   {e^{ - ik} } & {e^{ - ik} }  \\
   {e^{  ik}} & {-e^{  ik}}  \\
\end{array}} \right).
\end{equation}
Accordingly, all $C_i$ given in Eq. (\ref{C-position}) can be
rewritten as follows
\begin{eqnarray}\label{C-position-respectToU}
C_1&=&\sqrt {1 - p} U\left(k\right)\\\nonumber C_2&=&\sqrt{\frac
{p}{2}} e^{-ik} U\left(k\right)\\\nonumber C_3&=&\sqrt{\frac {p}{2}}
e^{ik} U\left(k\right).
\end{eqnarray}
We can , therefore, write
\begin{eqnarray}\label{L_rep_U}
  \mathcal{L}_{k,k'} \rho_0&=& \sum\limits_n{C_n\left(k\right) {\rho_0} C_n^\dag\left(k'\right)}\\
&=&\left[1-p+\frac{p}{2}\left(
e^{i\left(k'-k\right)}+e^{-i\left(k'-k\right)}\right)\right]\mathcal{W}_{k,k'}\rho_0\nonumber.
\end{eqnarray}
Putting this into Eq. (\ref{probability(x,t)}) and using  Eq.
(\ref{probability_0}), we find the following form for  the
probability distribution of decoherent QW
\begin{eqnarray}\label{decoherent_probability}
P\left(x,t\right)=\sum\limits^t_{n=0}\sum\limits^n_{m=0}&&
\left({\begin{array}{c} t\\n
\end{array}}\right)\left({\begin{array}{c} n\\m\end{array}}\right)
\times \\\nonumber
&&\left(1-p\right)^{t-n}\left(\frac{p}{2}\right)^n
P_0\left(x+2m-n,t\right),
\end{eqnarray}
where $\left({\begin{array}{c} r\\s
\end{array}}\right)=\frac{r!}{s!(r-s)!}$ is the binomial coefficient.

\begin{figure}
\centering
\includegraphics[width=9 cm]{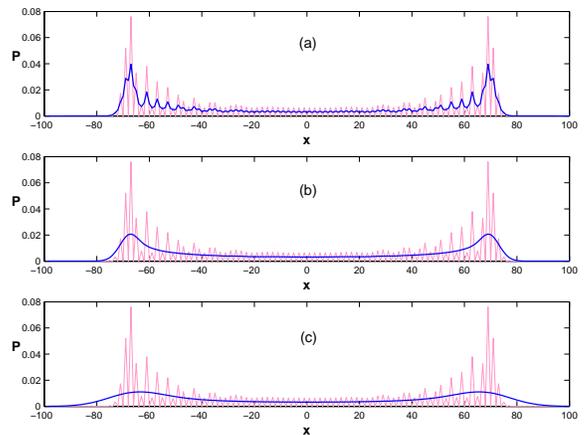}
\caption{(Color online) Probability distribution of the position
only decoherence (thick curve) for (a) $p=0.01$, (b) $p=0.1$ and (c)
$p=0.9$. The thin curve is the probability distribution of the
coherent QW, which is plotted for all sites and allows us to compare
all site occupations in the position decoherent QW with odd (even)
site occupations in the coherent case} \label{Fig:prob-dist}
\end{figure}

\begin{figure}
\centering
\includegraphics[width=9 cm]{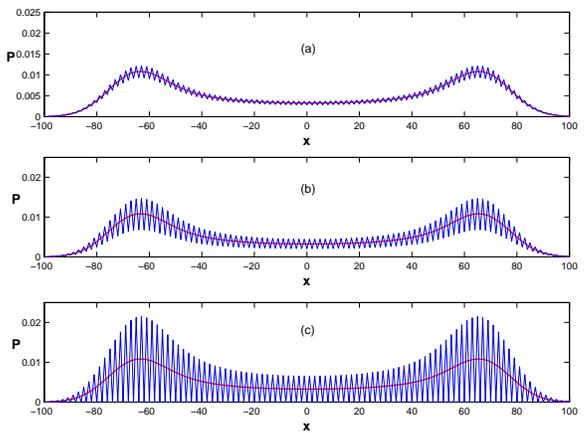}
\caption{(Color online) Probability distribution for strong values
of noise (a) $p=0.99$, (b) $p=0.995$ and (c) $p=1$. The thick and
smooth curve is, in all cases, the probability distributions for
$p=0.97$ which is approximately critical point in which the behavior
of the probability distribution changes from smoothness to rough
distribution by growing $p$.} \label{Fig:prob-dist-large-p}
\end{figure}

Equation (\ref{decoherent_probability})  shows that, after $t$
steps of decoherent walking, the probability of finding the walker
at site $x$  can be written as a linear combination of the
probabilities of the corresponding coherent walking, but at some
other sites ranging from $x-t$ to $x+t$. The equation can be also
used in order to prove the smooth property of the distribution and all sites occupation property. A comparison of this formula and
the numerical simulation shows that there are an excellent
consistent between them. Figure (\ref{Fig:prob-dist}) shows the
probability distribution in terms of $x$, in the presence of
position decoherence ( tunneling noise ) with different noise
strength.
It is worth to mention that although $P_0\left(x,t\right)=0$ for odd
(even) sites after even (odd) steps, but it is evident from Eq.
(\ref{decoherent_probability}) that the probability
$P\left(x,t\right)$ of finding the walker in all sites $x$ and
$0<p<1$ is nonzero (see Fig. (\ref{Fig:prob-dist})). In the totally
decoherent case, i.e. $p=1$, only the terms with $n=t$ have nonzero
contribution in the summation  of Eq.
(\ref{decoherent_probability}), leads therefore to
\begin{eqnarray}\label{decoherent_probability_p=1}
P\left(x,t\right)\Big|_{p = 1}=\frac{1}{2^t}\sum\limits^t_{m=0}
{\left({\begin{array}{c} t\\m\end{array}}\right)
P_0\left(x+2m-t,t\right)}
\end{eqnarray}
This equation shows that in the case $p=1$, unlike the coherent case
$p=0$, only even sites are occupied for any $t$. The reason is that
if $t=2k$, then  only for $x=2l$ we have
$P\left(x+2m-t,t\right)=P\left(2s,2k\right)\ne 0$ with $s=l+m-k$. On
the other hand, if $t=2k+1$, again only for $x=2l$, we have
$P\left(x+2m-t,t\right)=P\left(2s-1,2k+1\right)\ne 0$. This means
that the walker influenced by full tunneling noise, can not be found
in odd position after any number of steps. Numerical calculations
show that for $ p \gtrapprox 0.97 $ the probability distribution
loses it's smoothness and converges to the case $p=1$ in which only
half of the sites, namely even sites, are occupied. Figure
(\ref{Fig:prob-dist-large-p}) shows the probability distribution for
some large values of $p$.
It is evident from Figs. (\ref{Fig:prob-dist}) and
(\ref{Fig:prob-dist-large-p}) that for  very weak and very strong
noises the probability distribution is rough,  but for the other
strengths of noise the tunneling effect can smooths the probability
distribution. In order to quantify the smoothness of the probability
distribution in terms of $p$, we use total variation of the
probability distribution as a measure of roughness of distribution,
defined by \cite{Rudin}
\begin{equation}\label{total-variation}
    {\Var}^b_a\left( f \right) = \mathop {\sup }\limits_{S \in \mathcal{S}}
    \sum\limits_{i = 0}^{n_S - 1} {\left| {f\left( {x_{i + 1} } \right) - f\left( {x_i } \right)}
    \right|},
\end{equation}
where the supremum is taken over the set $\mathcal{S} = \left\{ {S =
\left\{ {x_0 ,...,x_{n_S } } \right\}|, S\;\textrm{is a partition
of}\;\left[ {a,b} \right]} \right\}$.  This definition is calculated
in the interval of all possible sites and is a relevant measure for
quantifying roughness of the probability distribution. In our
problem we have
\begin{equation}\label{roughness}
    {\Var}^t_{-t}\left( P \right) = \sum\limits_{i = -t}^{t - 1} {\left| {P\left( i + 1,t \right) - P\left( i,t \right)} \right|}.
\end{equation}
Note that in interval $\left[-t,t\right]$, the set
$S=\left\{-t,...,t\right\}$ includes all sites and we have discreet
values for probability, so supremum occurs automatically.

\begin{figure}
\centering
\includegraphics[width=9 cm]{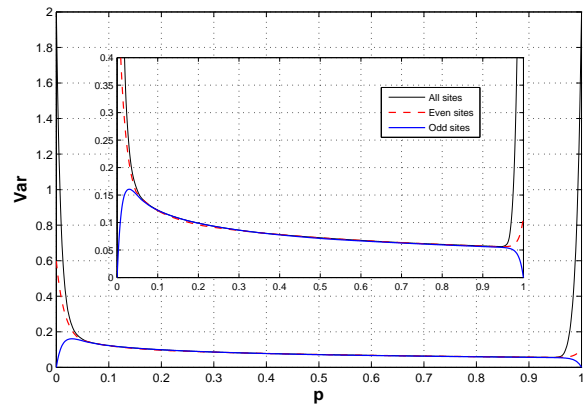}
\caption{(Color online) Total variation of probability distribution
of QW  in the presence of position decoherence versus p. The upper
panel is close up of curve for presentation of more details }
\label{Fig:Var}
\end{figure}

Figure (\ref{Fig:Var}) shows the total variation of the probability
distribution with respect to $p$. As we expect, for small value of
$p$, the total variation is large, which means that the probability
distribution is not different from the coherent case yet. But, on
the other hand,  the total variation decays very fast and
approximately remains flat, except for $p$ near to 1 which in this
case, as it is clear from figure, the total variation increases
critically. It means that the noise applied on the position of the
walker changes the probability distribution such that one can find
the walker in the neighbor sites with approximately equal
probability. But for large values of $p$, a gap between
probabilities  of neighbor sites occurs which is very sensitive by
applying a bit of noise.

At the end of this section, we would like to emphasis that the
expression in Eq. (\ref{decoherent_probability}) is independent of
the type of walking. It means that for any type of 1DQW with known
coherent probability distribution, one can use this expression to
find the effects of position noise on it.  Several types of 1DQW
such as biased Hadamard walk \cite{Carneiro}, QW with SU(2) coin
operator \cite{Chandrashekar_SU(2)QW}, 1DQW with entangled coin
\cite{Venegas}, many coins QW \cite{Brun-ManyCoin} and
so on have been introduced and well studied. Some aspects of
specific types of 1DQW can be combined with the smoothness property
of the position decoherence and opens new useful features in the
study of QW. For example, Kendon et al. \cite{KT03} have shown that
the weak decoherence, both on the position and on the coin
subspaces, can produces flat probability distribution where are
useful in quantum information and computation procedures. In fact,
in their work the coin decoherence tends to change the quantum
probability to classic one  (binomial distribution around the
origin) and the position decoherence smooths the distribution. It is
clear that applying noise on the coin subspace decreases the
quantumness of the system, so their work has made in the cost of
decrease of the quantum features of the system such as the speed of
spread and the variance. One of the promising interesting  aspect of
the current work is that we could smooth the probability
distribution without the lost of the quantum features. For
example, we can use the many coins \cite{Brun-ManyCoin} or the 1DQW with
entangled coins \cite{Venegas}, both having the  probability
distributions with some peaks, and smooth the
probability distribution by tuning the noise on the position
subspace.

\section{summary and conclusions}
By using the analytical expressions for the moments of distribution,
given recently in  \cite{annabestani-deco},  we have investigated
the tunneling effect in the one-dimensional quantum walk. Our
analytical calculation shows that the shifting noise (tunneling
effect) on the position subspace of the quantum walk does not fade
the quantum behavior of the system. We have brought two witnesses
for our claim,  which behave completely different when the coin
subspace is subjected to the decoherence. First, we have shown  that
the quadratic dependency of the variance on the time has never
vanished, even for mixed initial state. Second,   we have calculated
the entanglement between the coin and the position and have shown
that unlike the coin only decoherence, in which the entanglement
goes to zero very fast by increasing the strength of the noise, the
entanglement in the position only decoherence (tunneling effect)
converges to a  significant value, even for maximum rate of noise.
Furthermore, we have derived an exact expression for the probability
distribution of the QW, in the presence of tunneling effect, in
terms of the corresponding coherent probability distributions. Our
results have shown that the effect of the tunneling noise on the  QW
is the smoothness of the probability distribution, such that this
smoothness  occurs for all noise strength $p$, except for $p$ near
to 1 for which the smoothness is broken. Moreover, we found that although for $p$ near to 1, the probability distributions of the even and odd sites are separate from each other, but even for this case the probabilities
distribution of the odd and even sites are smooth individually (see
Fig. \ref{Fig:Var}). Another interesting result which we have found
is that,  for maximum rate of noise, i.e. $p=1$, only the even sites
are occupied despite of the number of steps.

Accordingly, in the all experimental realization of the QW where the
the quantum behavior, such as the speed of spreading or the
entanglement, plays an important rule, one does not need be worry
about the noise on the position subspace, because  this type of
decoherency does not fade the quantumness of the system, but the
significant classicality will happen whenever the coin is subjected
to the decoherence. The present paper shows that the position only
noise not only does not make the transition from quantum to classic
but also it can smooths the probability distribution, a task  which
is useful in quantum information and quantum computation process.

Since the smoothness of the probability distribution and the exact
expression  of Eq. (\ref{probability(x,t)}) are independent from the
type of the QW, it can be used with other aspects of different type
of QW in order to invent useful process in quantum information and
quantum computation. As an example, many coins QW introduced by Brun
et al \cite{Brun-ManyCoin},  persists the quantum behavior of
coherent 1DQW for $M<t$, where $M$ is the number of coins. Also this
type of QW has many picks in the probability distribution, which can
be smoothed by applying the proper noise on the position subspace.
As an another example, we could mention the quantum walk on the line
with entangled coins \cite{Venegas}. In this type
of the QW the probability distribution has a pick around the origin
which can be smoothed by using the tunneling noise without worry
about the classicality of the system. Therefore the results of this
paper and other aspects of the 1DQWs  can be combined together and
open a  new window in order to using properties of QW in quantum
information and quantum computation processes.

\acknowledgements{M.A. thanks  M. Amini and G. Abal for valuable
discussions.}

\bibliography{QW_correctedByAPSforma}

\end{document}